\documentclass[aps,prl,floatfix,twocolumn,showpacs,preprintnumbers,amsmath,amssymb]{revtex4}

\usepackage{epsfig}
\usepackage{cancel}
\usepackage{hyperref}
\usepackage[usenames]{color}

\begin{document}

\title{Collisional stability of fermionic Feshbach molecules}
\author{J. J. Zirbel}
 \email{zirbel@jilau1.colorado.edu}
\author{K.-K. Ni}
\author{S. Ospelkaus}
\author{J. P. D'Incao}
\author{C. E. Wieman}
\author{J. Ye}
\author{D. S. Jin}

\affiliation{JILA, Quantum Physics Division, National Institute of Standards and Technology and the Department of Physics, University of Colorado, Boulder, CO 80309-0440, USA}

\begin{abstract}
Using a Feshbach resonance, we create ultracold fermionic molecules starting from a Bose-Fermi atom gas mixture. The resulting mixture of atoms and weakly bound molecules provides a rich system for studying few-body collisions because of the variety of atomic collision partners for molecules; either bosonic, fermionic, or distinguishable atoms.  Inelastic loss of the molecules near the Feshbach resonance is dramatically affected by the quantum statistics of the colliding particles and the scattering length.  In particular, we observe a molecule lifetime as long as 100\,ms near the Feshbach resonance.
\end{abstract}

%36.90.+f, 03.75.Ss, 03.75.Nt
\maketitle

Ultracold Bose and Fermi gases are rich systems for the study of few-body collisional properties. These collisional properties can dramatically impact the many-body behavior of ultracold gases.   In particular, recent studies using magnetic-field controlled two-body scattering resonances called Feshbach resonances have demonstrated efficient creation of diatomic molecules from strongly interacting atoms \cite{Kohler2006}. In most of these studies, the molecules formed are composite bosons, which are particularly interesting when the original atoms are fermions. In this case, collisional decay of the molecules is strongly suppressed due to Fermi statistics\,\cite{Petrov2004a}.  When bosonic molecules are instead created in an atomic Bose gas, this suppression is not present and the molecules tend to have much shorter lifetimes \cite{Yurovsky1999}.

Fermionic molecules have recently been created in an ultracold Bose-Fermi gas mixture \cite{Ospelkaus2006}. Since these molecules and surrounding atoms were isolated in single wells of an optical lattice trap, the stability of these fermionic molecules against collisional decay has not been addressed experimentally and is  still an open question. In this letter, we present a systematic study of inelastic collisions starting from a Bose-Fermi mixture of atoms near a heteronuclear Feshbach resonance. Using the Feshbach resonance, we create ultracold heteronuclear molecules and study their inelastic decay. We find that the dominant decay process arises from atom-molecule collisions and observe the effect of both fermionic suppression and bosonic enhancement on the inelastic collisions.  The effects of quantum statistics arise because of indistinguishability of the colliding atom and one of the constituents of the molecule.  With appropriate atomic collision partners, we observe molecular lifetimes as long as 100\,ms. For large scattering length, $a$, experimental data are compared to predictions based on a numerical solution of the three-body Schr\"odinger equation \cite{Suno2002}.

Understanding the loss mechanisms in Bose-Fermi mixtures near heteronuclear Feshbach resonances is essential for experimental exploration of interesting many-body effects in these systems\,\cite{Efremov2002}.  Moreover, understanding and controlling these collisional loss processes is important for proposed experiments using heteronuclear Feshbach molecules as the starting point for the creation of ultracold polar molecules (\cite{doyle2004} and references therein).

Efficient molecule creation near a Feshbach resonance requires a gas near the quantum degenerate regime\,\cite{Hodby}.  We use fermionic $^{40}$K and bosonic $^{87}$Rb atoms, which we first trap and cool using standard magneto optical trap (MOT) techniques.  We then load $10^{9}$ Rb $|2,2\rangle$ atoms and $10^{7}$ K $|9/2,9/2\rangle$ atoms into a moving quadrupole magnetic trap\,\cite{Lewandowski2003} and transfer them to an ultra-low-pressure section of our vacuum chamber.  Here, we load a Ioffe-Pritchard type magnetic trap and use forced evaporation to cool the Rb gas, which in turn sympathetically cools the K gas. At the end of evaporation, $5\times 10^{6}$ Rb and $1\times 10^{6}$ K atoms at 3 $\mu$K are loaded into an optical dipole trap.  This trap is formed by a single focused laser beam with a $1/e^{2}$ radius of 50 $\mu$m and a wavelength of 1064 nm.  After loading the optical trap, we transfer Rb atoms into the $|1,1\rangle$ state and K atoms into the $|9/2,-7/2\rangle$ state using rf adiabatic rapid passage. Here, $|f,m_{f}\rangle$ denotes the hyperfine spin state with total atomic spin, $f$, and projection along the magnetic field direction, $m_{f}$.  We apply a magnetic field of 540\,G and evaporate both Rb and K gases to produce $3\times10^{5}$ Rb at $T/T_{\textrm{c}} \sim 1$ and $1\times10^{5}$ K at $T/T_{\textrm{F}} \sim 0.6$ in an optical trap.  The trapping frequencies are 136 Hz radially and 2.5 Hz axially for the Rb atoms.

To create molecules, we use a Feshbach resonance between Rb $|1,1\rangle$ and K $|9/2,-9/2\rangle$ atoms at $B_{0}=546.7$ G \cite{Inouye2004}.  We apply a transverse rf magnetic field to convert free atom pairs to molecules \cite{Ospelkaus2006}.  The strength of the rf field gives a $12\,\mu$s $\pi$-pulse on the atomic K Zeeman transition, $|9/2,-7/2\rangle\!\! \rightarrow \!\!|9/2,-9/2\rangle$.  To create molecules, we detune the rf frequency from the atomic transition by an amount corresponding to the binding energy of the Feshbach molecule. For example, at $545.97$ G, the frequency for maximum molecule creation is $80.23$ MHz. Here, we typically produce $1.5 \times 10^{4}$ molecules at a temperature of 150 nK and a density  of $\sim10^{12}/\rm cm^3$.

The molecules are imaged in absorption using light resonant with the K  $|9/2,-9/2\rangle\rightarrow|11/2,-11/2\rangle$ cycling transition at high magnetic field\,\cite{Ospelkaus2006, Zirbel2007c}. We measure the lifetime of the molecules in the weakly bound Feshbach state.  The decay mechanism for the Feshbach molecules is assumed to be caused by atom-molecule collisional vibrational quenching; these inelastic collisions produce more deeply bound molecules with enough kinetic energy to escape from the trap.  We characterize the decay of the cloud by its $1/e$ lifetime, $\tau$, and introduce a loss coefficient, $\beta$, which characterizes the strength of vibrational relaxation.  This coefficient is related to $\tau$ by $\beta=\frac{1}{n_{\textrm{A}} \tau}$, where $n_{\textrm{A}}$ is the initial average density of the dominant atomic collision partner, $n_{\textrm{A}}=\frac{1} {N} \int n(\textbf{r})^{2}d^{3}r$.

\begin{table}[t]
  \begin{tabular}{|c|c|}\hline
  Collision partners & Expected dependence for $a>0$ \\\hline\hline
 \textit{X+BF} & $\beta(a)  \propto a^{-1}$ \,\cite{D'Incao2007}    \\\hline
 \textit{B+BF} & $\beta(a)  \propto  P(a) \times a$ \,\cite{D'Incao_combined, Braaten2004} \\\hline
 \textit{F+BF} & $\beta(a) \propto a^{-3.12}$ \,\cite{D'Incao_combined}\\\hline
 \textit{B+B+F} & $K_{3}(a) \propto M(a) \times a^4$ \,\cite{D'Incao_combined}\\\hline
   \end{tabular}
   \caption{\label{tab:threeloss} . `\textit{BF}' refers to the Feshbach molecule, `\textit{B}' to the boson, `\textit{F}' to the fermion, and `\textit{X}' to a distinguishable K atom. $P(a)$ and $M(a)$ are log periodic functions in $a$ with modulations resulting from the formation of Efimov states. }
\end{table}

For Feshbach molecules, $\beta$ will depend on the magnetic-field detuning from the Feshbach resonance, $B-B_{0}$.  This detuning is related to the heteronuclear scattering length, which is given by $a=a_{bg}(1-\frac{W}{B-B_{0}})$.  The background scattering length is $a_{bg}=-185 \,a_{0}$, where $a_{0}$ is the Bohr radius, and the width of the resonance is $W=-2.9$\,G \cite{Ferlaino2006}.  In the limit of zero temperature and large $a$ ($a\gg r_{\mathrm{vdW}}$, where $r_{\rm vdW} = 72 a_0$ is the van der Waals length \cite{Kohler2006}), $\beta$ is predicted to have a universal power law dependence on $a$\,\,\cite{D'Incao_combined}.  The power laws are strongly dependent on the quantum statistics of particular collision partners (see table I).  While our measurements are not fully in the limit of $T=0$ and large $a$, we are able to observe the strong influence of the quantum statistics of the particles  involved in an inelastic atom-molecule collision. For large $a$, we compare our experimental data with finite temperature numerical calculations obtained by solving the Schr\"odinger equation in the hyperspherical adiabatic representation \cite{Suno2002}. We have assumed a simple finite range potential model for the interatomic potentials $v(r)=-D\,\mathrm{sech}^2(r/r_{\rm vdW})$, where $r$ is the interatomic distance and $D$ is adjusted to produce the desired variations in $a$\,\cite{note1}. For each comparison with the experimental data, the numerically calculated inelastic rate is scaled by a multiplicative factor. Collisional dissociation occurs at finite temperature ($T\approx 150$ nK) when the binding energy of the molecules is comparable to the collision energy, ($a\gtrsim 4500\,a_0$) \cite{Braaten2006, D'Incao_combined}.  We restrict our measurements in a regime where collisional dissociation is not a problem.

In measuring the loss rate $\beta$, one would ideally like to isolate one particular inelastic collisional process.  Experimentally, we can control the internal state of the Rb and K atoms using applied rf fields.  Combining this with resonant light pulses, we control the atom densities through selective removal from the trap.  We first consider the case of distinguishable atoms colliding with molecules, where the quantum statistics of the colliding particles is not important.  The distinguishable atom is chosen to be K in the $|9/2,-7/2\rangle$ state.  This state is stable against inelastic two-body collisions with either K $|9/2,-9/2\rangle$ or Rb $|1,1\rangle$ atoms.  A nearly pure mixture of K $|9/2,-7/2\rangle$ atoms and Feshbach molecules is prepared by removing excess Rb atoms from the trap after creating molecules.  Using a combination of microwaves to drive Rb atoms to the $|2,2\rangle$ state and resonant light on the cycling transition, we remove 99\% of the Rb atoms and leave the K $|9/2,-7/2\rangle$ atoms and molecules essentially unperturbed.

    \begin{figure}[t]
    \includegraphics[width=0.8\columnwidth]{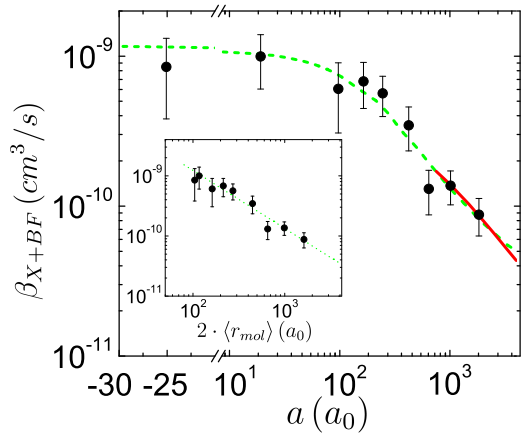}
    \caption{\label{fig:DisK}  \textbf{(color online)} Molecule loss coefficient $\beta$ for collisions with distinguishable atoms (K in the $|9/2,-7/2\rangle$ state) as a function of the heteronuclear scattering length.  For large $a$, we observe $\beta$ to decrease with increasing scattering length in agreement with numerical calculations (solid curve).    We observe $\beta$ saturating for small $a$, ($a\sim 100\,a_0$) where the molecular size, $\langle r_{mol}\rangle$, no longer has a strong dependence on $a$. Inset: $\beta$  vs.  molecule size  $\langle r_{mol}\rangle$ as extracted from  the measured binding energy of the Feshbach molecules\,\cite{Zirbel2007b}. The dashed curve is a power law fit $\beta_{\rm fit}\propto  \langle r_{mol}\rangle^p$ to the experimental data; we obtain $p=-0.97\pm 0.16$. }
    \end{figure}

Figure \ref{fig:DisK} shows the measured $\beta$ as a function of the heteronuclear scattering length along with numerical calculations for this case.  As predicted \cite{D'Incao2007}, the loss rate, $\beta$, decreases near the Feshbach resonance (increasing $a$).  Even though the quantum statistics for this case is not important, $\beta$ is expected to be suppressed as $a^{-1}$ as a result of the repulsive character of the effective atom-molecule interaction.  The suppression for smaller scattering lengths, but reasonably large molecule sizes, can be qualitatively explained by an intuitive argument involving poor wavefunction overlap with deeply bound molecular states. For our smallest scattering lengths, we find that $\beta$ stops changing with decreasing $a$.  Here, the universal prediction for the Feshbach molecule size, $a/2$, is no longer valid.  To account for this effect we can instead plot $\beta$ versus an improved estimate of the molecular size, $\langle r_{mol} \rangle$, which takes into account both closed and open channel contributions \cite{Zirbel2007b}. Fitting the data in Fig. \ref{fig:DisK} (inset) we find a power law of $-0.97\pm0.16$.

    \begin{figure}[t]
    \includegraphics[width=0.8 \columnwidth]{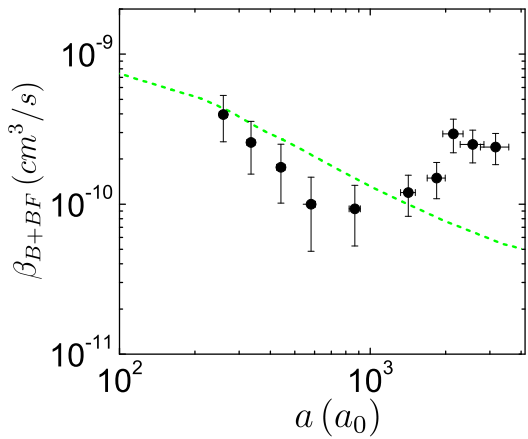}
    \caption{\label{fig:Boson} \textbf{(color online)}
    Molecule loss coefficient for collisions with indistinguishable bosons  versus heteronuclear scattering length.  For large $a$, $\beta$ increases with increasing $a$. In this regime the molecular loss is enhanced compared to loss due to molecule collisions with distinguishable atoms (dashed curve and Fig. 1). For small $a$, $\beta$ increases with decreasing scattering length (see text).
    }
    \end{figure}

Next, we present measurements of molecule collisions with bosonic atoms.  Here we measure the inelastic decay rates of Feshbach molecules when both Rb $|1,1\rangle$ atoms and K $|9/2,-7/2\rangle$ atoms are present.  The data was taken with the Rb density about 3x larger than the K density. The contribution of collisions with distinguishable K is known (Fig. \ref{fig:DisK}) and subtracted from the data to extract the loss coefficient, $\beta$, for collisions with Rb only.

Figure \ref{fig:Boson} shows a plot of $\beta$ as a function of the heteronuclear scattering length, $a$.  For large $a$, $a \gtrsim 1000\,a_{0} $, we observe a distinct enhancement of $\beta$ that increases with increasing $a$. This agrees with the expectation that Bose statistics should enhance the molecule decay rates  due to the attractive character of the effective atom-molecule interaction\,\cite{D'Incao_combined}.  In addition,
 Efimov three-body bound states might induce enhanced resonant losses due to trimer formation\,\cite{Kraemer2006, Braaten2006}. The present data do not rule out Efimov states. However, we also do not observe a clear signature of Efimov states for the range of scattering lengths and temperatures carried out in this work.

 The measured molecular loss coefficients increase for decreasing $a$ for $a \lesssim 1000\, a_{0}$.  This trend is reminiscent of collisions with distinguishable atoms. One possible explanation is that the increasing closed channel fraction of the Feshbach molecule far from resonance makes the colliding free atom effectively distinguishable from the molecule's constituent atoms, and hence decreases the influence of bosonic enhancement.

    \begin{figure}[t]
    \includegraphics[width=0.8\columnwidth]{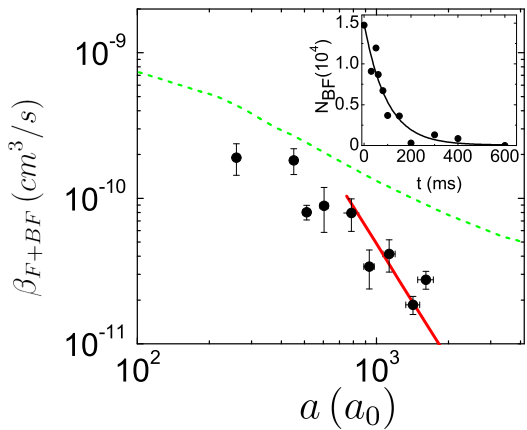}
    \caption{\label{fig:Fermi} \textbf{(color online)} Molecule loss coefficient $\beta$ for collisions with indistinguishable fermions.  For large $a$ we observe  loss rates  scaling as $a^{-1.6\pm0.2}$. The power law dependence is consistent with numerical calculations for this scattering length regime (solid curve). In addition, the molecular loss is suppressed compared to inelastic decay of  molecules due to collisions with distinguishable atoms (dashed curve and Fig. 1). Inset: Molecule decay at $a\approx1300\,a_0$. The $1/e$-lifetime of the molecular cloud is $\tau=100\pm 20\,\rm ms$ }
    \end{figure}

To study collisions between molecules and fermionic atoms, we prepare a nearly pure mixture of K $|9/2,-9/2\rangle$ atoms and Feshbach molecules using the following procedure: after creation of the molecules, we remove Rb atoms from the trap and drive the remaining K atoms to the $|9/2,-9/2\rangle$ state with an rf $\pi$-pulse on the atomic transition.  This rf pulse does not affect the molecules since its frequency is detuned by about five linewidths. After waiting a variable time for the molecule decay measurement, we drive the K atoms back into the dark  $|9/2,-7/2\rangle$ state and image the remaining molecules.

Experimental results are shown in Figure \ref{fig:Fermi}.  These data agree with the theoretical expectation that the decay is suppressed with increasing $a$.  For our measurements at finite $a$, we observe $\beta$ scaling as $a^{-1.6\pm0.2}$. This trend qualitatively agrees with numerical calculations (solid curve in Fig. \ref{fig:Fermi}). The suppression of molecular decay allows for molecule lifetimes as long as $100$\,ms near the resonance (see inset of Fig.\,\ref{fig:Fermi}).  We note that the calculated $\beta$ eventually scales as $a^{-3.12}$ for larger $a$\,\cite{D'Incao_combined}.

For the fermionic case (Fig. \ref{fig:Fermi}), we observe a suppression of $\beta_{\rm F+BF}$ when compared to the decay rates due to collisions with distinguishable atoms (Fig. \ref{fig:DisK}). In contrast, for the bosonic case (Fig. \ref{fig:Boson}), we observe $\beta_{\rm B+BF}$ to be larger than the distinguishable atom case near the resonance.  The generality of this behavior is an open question because $\beta$ depends on short-range physics that will be different in other systems.

Molecule-molecule collisions \cite{Knoop2007} might also contribute to the inelastic loss of molecules, but these are expected to be suppressed due to the fermionic character of the KRb molecules.  In the analysis, we have assumed no molecular decay due to molecule-molecule collisions.  From our longest molecule lifetimes, an upper limit on the molecule-molecule loss coefficient at a scattering length of $a=1500\,a_{0}$ is $\beta_{\textrm{BF+BF}} \leq 4\times10^{-11}$cm$^{3}/$s.

    \begin{figure}[t]
    \includegraphics[width=0.7\columnwidth]{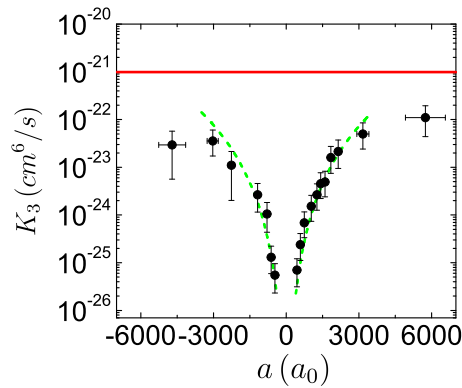}
    \caption{\label{fig:ThreeBody} \textbf{(color online)} Three-body recombination near the heteronuclear Feshbach resonance.  The measured $K_{3}$ varies as $a^{4}$ and saturates at large $a$ due to the unitary limit.  The dashed curve is an $a^4$ fit to the data for $|a|<4000\,a_{0}$. The solid line is the unitarity limit at 150\,nK.  }
    \end{figure}

We finish the three-body studies with measurements of three-body recombination rates for atoms in a Bose-Fermi mixture near a heteronuclear Feshbach resonance.  We extract the three-body loss coefficient, $K_{3}$ as defined by \cite{note2}, by analyzing measurements of the time-dependent loss of the atomic species as a function of $a$ on both sides of the resonance (Fig.\,\ref{fig:ThreeBody}).  We find that near the Feshbach resonance, the loss rate $K_3$ is enhanced by many orders of magnitude. It follows the expected power law dependence  $K_{3} \propto |a|^{4}$\,\cite{D'Incao_combined} before saturating near the resonance due to the onset of unitarity for scattering lengths $|a|>4000\,a_0$ \cite{D'Incao2004}.  Fitting the data to $K_{3}=\eta \times (a/a_{0})^{4}$ on both sides of the resonance, we find $\eta=(9.4\pm1.2)\times 10^{-37}$ cm$^{6}$/s.

In conclusion, we have created weakly bound fermionic molecules near a heteronuclear Feshbach resonance and studied inelastic, atom-molecule loss processes. We observe that the quantum statistics of the particles strongly affects vibrational quenching rates of the molecules near the Feshbach resonance.  Our data are consistent with the expectation that collisions between identical fermionic molecules are suppressed.  This allows for lifetimes of the fermionic heteronuclear molecules as long as 100\,ms, which is at least ten times longer than required for coherent manipulation steps and opens the possibility of production of ultracold polar molecules.  However, when bosonic atoms are available to collide with the molecules, we observe enhanced decay rates.  Therefore, studies of the many-body behavior of a Bose-Fermi mixture near a Feshbach resonance may prove challenging.

We thank T. Nicholson, A. Wilson, and B. Neyenhuis for experimental assistance; A. Pe'er and C. Greene for useful discussions; B. Esry for access to his three-body code.  K.-K. Ni and J. P. D. acknowledge support from the NSF and S. O. from the A.-v.\,Humboldt Foundation.  This work was supported by NSF, NIST, NERSC, and the W. M. Keck Foundation.

%\bibliography{moleculep}

\begin{thebibliography}{10}
\bibitem{Zirbel2007c}
For weakly bound heteronuclear Feshbach molecules, the atomic K cycling
  transition is detuned only by a few kHz to MHz from the molecular transition
  at high magnetic field. One photon dissociates the molecule and subsequent
  photons scatter on the resulting K atoms.
\bibitem{note1}
We limit our comparison between experiment and theory to a scattering length
  regime where the finite-range model can reproduce experimentally observed
  binding energies to within 20\%. The cut-off occurs at $a=700\,a_0$.
\bibitem{note2}
$ \frac{dn_{\textrm{K}}} {dt} = -K_{3}\frac{n_{\textrm{Rb}}^{2}n_{\textrm{K}}}{2!}$  where $n_{\textrm{K}}$ and $n_{\textrm{Rb}}$ are the K and Rb densities, respectively.
\end{thebibliography}

\end{document}